# Trial and Return Option Strategy in Omnichannel Retailing

Yasuyuki Kusuda[a]*

[a]*Faculty of Economics, Tokyo International University, 4-42-31 Higashi-Ikebukuro, Toshima, Tokyo, 170-0013, Japan*

*Corresponding author: Faculty of Economics, Tokyo International University, *E-mail address:* ykusuda@tiu.ac.jp (Y. Kusuda)

**Abstract**

This study examines the dynamics of customer behavior with trial and return options in omnichannel retailing, where retailers face challenges in integrating physical and online stores. Recently, major retailers have begun offering customers the option of trying eligible items for a set period and returning unwanted products free of charge. However, existing research has not fully explored the temporal dynamics of customer return behaviors. This study investigates how temporal dynamics affect customer return behaviors and decision-making during trial periods. Using a theoretical game structure framework, this study explores customer decision patterns regarding store visits, product trials, and returns, while examining the strategic role of store clerks in encouraging product trials. The findings suggest that retailers can maximize profit through trial and return options when product fit probability is low, emphasizing the importance of maintaining complementary physical and online channels. We also found that store clerks play a critical role in encouraging customers to try misfit products. The results further reveal that return cost coverage policies may not significantly impact customer behavior or retailer profit.

## 1. Introduction

As the retail landscape continues to evolve towards an omnichannel approach, retailers are confronting increasingly sophisticated challenges in seamlessly integrating their physical and online channels. The emergence and growing popularity of "try-before-you-buy" programs and customer-friendly return policies have introduced a new layer of complexity into the delicate balance between enhancing customer satisfaction and maintaining operational efficiency. In this evolving ecosystem, physical retail locations have undergone remarkable transformation, simultaneously functioning as traditional points of sale while also serving as crucial service hubs for online purchases. Online platforms must maintain intricate coordination with traditional brick-and-mortar stores to deliver a unified, friction-free shopping experience that modern customers expect.

Major retailers have begun offering customers the option of trying eligible items for a set period and returning unwanted products free of charge. For example, *Amazon Prime Try Before You Buy* lets Prime members try eligible clothing, shoes, and accessories at home for up to seven days. Members pay only for the items they keep and can return unwanted items at no cost. Apple's standard return policy grants customers 14 days to return most items after receipt. For online purchases, customers can initiate returns online and choose between shipping items back or returning them to an Apple Store. The company offers free return shipping for most products.

Recent research on omnichannel firms' return policies has focused on refund strategies and how easily customers can access stores for returns. For example, Nageswaran et al. (2020) analyzed why omnichannel firms offer generous refunds, contrasting them with single-channel firms. Their findings showed that omnichannel firms should consider improving customers' accessibility to physical stores when deciding on full refund policies. Opening new stores can boost revenue because customers who visit stores are less likely to make returns. In contrast, Yang et al. (2023) examined how buy-online-and-return-to-store (BORS) affects customer acquisition. They argued that although BORS encourages online purchases, it does not necessarily increase retailer profits or store traffic. Their research suggested that these changes in customer behavior may not benefit retailers.

However, existing research has not fully explored the temporal dynamics of customer return behaviors. The introduction of product trial periods, as opposed to traditional immediate return policies, introduces a crucial temporal dimension that can significantly impact customer decision-making. When customers have extended periods to evaluate products, their return decisions are influenced by their developing tolerance and familiarity with the items over time, which may lead to decreased return rates compared with immediate return scenarios. During extended trial periods, customers often face psychological barriers when contemplating returns, particularly when they must absorb substantial return costs. The presence of these return-related expenses can serve as a powerful deterrent against returns, ultimately working in the retailer's favor when the profit margins of the products are considerable. In this context, the traditional advantages associated with full refund policies may become secondary considerations in the overall retail strategy.

To explore the dynamics of customer behavior, a theoretical analysis using an extensive game structure was conducted. In this framework, customers make several key decisions that shape their shopping journey: whether to visit a physical store, try the product even if it does not fit their preferences, and return it after the trial period. Based on these customer decisions, the retailer aims to maximize profits by coordinating retail channels and offering trial options. When the trial option proves beneficial, the retailer must leverage the store clerks—available only in physical stores—to encourage customers to use this option. This strategic deployment of human resources in physical stores provides an additional compelling rationale for implementing and maintaining omnichannel retail systems. The trial option also affects how the retailer sets the product prices. The retailer must lower prices to encourage customers to try the product and keep it after the trial period. Setting the optimal price is complex, as maximizing profit may involve a strategy in which *some* customers keep the product while others return it, rather than aiming for *all* customers to keep it.

Therefore, the following research questions were examined in this study:

1. How can omnichannel retailers effectively coordinate retail prices while offering product trial options to customers?
2. What is the role of physical store clerks in an omnichannel system and when do they become crucial?

3. Is it profitable for retailers to cover return costs for customers?

The main findings of this study are as follows. First, retailers can maximize profits by offering trial and return options when both return costs and the probability of product misfit are low. Otherwise, it would be more beneficial for retailers not to offer this option, as they would need to lower product prices, resulting in reduced profits. Customers who try a product that does not fit their preferences must decide whether to keep or return it. When return costs are high, customers tend to keep the product. However, when the return costs are low, some customers choose to return it. Therefore, retailers should maintain two complementary channels: physical stores for product trials and online stores to provide alternatives when products do not fit. Second, retailers should maintain an omnichannel system that integrates both physical and online stores. Physical stores are especially valuable because store clerks can encourage customers to try products they may initially dismiss. This human element in physical stores provides a unique advantage in the omnichannel system by helping customers overcome their initial hesitation about unfamiliar products, a key benefit of implementing an omnichannel strategy. Third, in our model, when retailers cover the return costs for customers who try and return misfit products after the trial period, this coverage policy does not affect customer behavior or retailer profit. When a retailer announces full coverage of return costs, customers know that they can return the products during the trial period without paying return fees, which encourages them to try the products.

This study contributes to the existing omnichannel retailing discussion by demonstrating that trial and return options provide another rationale for omnichannel systems. First, it suggests that retailers must maintain two types of retail channels: physical stores, where customers can examine products in person, and online stores, which offer alternative products that do not require physical inspection. Second, it suggests that, in an omnichannel system, store clerks play a vital role in recommending trial options to customers who find products that do not meet their needs. These clerks can guide customers to alternative products available online if they decide to return their initial purchases.

The uniqueness of this study lies in examining dynamic customer behavior regarding trial and return options using an extensive game structure framework. Drawing on Gu and Tayi's (2017) concepts of *fit probability* and *tolerance levels* of customers, we analyze how customers decide whether to keep or return misfit products during trial periods. We demonstrate how customer behavior influences retailer profits, which helps establish optimal omnichannel retailing systems.

The remainder of this paper is organized as follows. Section 2 reviews the relevant literature. Section 3 presents the theoretical model for analyzing optimal customer behavior and retailer profits. Section 4 discusses the findings of the model analysis and Section 5 concludes the paper.

## 2. Literature Review

Return behavior research falls into two main streams: single-channel and omnichannel retailing. This section reviews return behavior studies in omnichannel retailing. Research on omnichannel return policies typically examines the relationship between returns and other aspects of retail. According to Nageswaran et al. (2020) and Yang et al. (2023), these aspects include return windows, return policies, money-back guarantees, refund policies, and customer behavior.

Sharma and Dutta (2023) provided a comprehensive survey of omnichannel-related literature, classifying it into five clusters based on keyword analysis: supply chain and operations, customer behavior, technological innovation, channel integration, and omnichannel strategy. Among these, supply chain and operations and omnichannel strategy are the most relevant to return policies. Specifically, return window considerations fall under supply chain and operations, while other return-related topics align with the omnichannel strategy.

de Borba et al. (2021) conducted a systematic literature review of return management challenges in omnichannel retail and identified 43 barriers such as high investments, restocking costs, and communication issues. Their research establishes a theoretical framework for understanding reverse logistics in omnichannel retail, while noting the field's emerging nature. Their findings offer guidance to retail and supply chain managers in handling omnichannel complexity, especially after-sales return management. Chen and Chen (2017) proposed a dual-channel model for retailers to optimize online channel introduction and return policy structures. They recommended money-back guarantees when the net salvage value is positive and suggested independent channel pricing under uniform pricing schemes. Zhang et al. (2017) investigated how return policies, particularly return windows and refund amounts, could signal product and service quality to online shoppers with incomplete product information. They found that long return windows and full refunds indicate high service quality and boost purchase intention, return windows do not signal product quality, and the impact of return depth varies with service quality. Hua et al. (2017) analyzed retailers' shipping and return charge decisions under no-reason return policies. Their theoretical model showed that retailers achieve better results by making shipping and return charge decisions simultaneously, rather than separately, often by pairing free shipping with higher return charges and shipping fees with lower return charges.

The present study focuses on return policies in relation to channel strategies and customer behavior. Recent studies have explored several aspects of this topic. Mandal et al. (2021) analyzed three omnichannel retail configurations—online-only, experience-in-store-and-buy-online, and BORS—addressing the high e-commerce return rates caused by customers' inability to examine products before purchase. Their study revealed that the optimal channel strategy varies by product attribute: premium personalized items benefit from showrooms or brick-and-mortar stores, whereas low-value personalized products are better suited to exclusive showrooms. Nageswaran et al. (2020) investigated how omnichannel retailers should design return policies by considering factors such as salvage partnerships and store networks. They found that full refunds are most effective for firms with strong salvage partners or store-based customers, whereas retailers with extensive store networks may benefit from online return fees that encourage in-store returns. Furthermore, Yang et al. (2023) studied how BORS policies impact retail operations. While BORS can attract new customers and lower in-store inventory requirements, its effectiveness depends on low return penalties and rates and may reduce profitability when resalable returns are high.

Regarding trial options with returns, Jena (2022) examined how test-in-store-and-buy-online (TSBO) retailing compares with online retail when considering different bundling strategies, focusing on a supply chain with two manufacturers and one retailer. The findings show that TSBO retailing generates the highest profit when the second manufacturer integrates with the online retailer, although this advantage decreases as bundling costs increase above a certain threshold. Jena and Meena (2022a) analyzed how the TSBO retail strategy influences supply chain profits and manufacturer competition in an omnichannel setting. They concluded that TSBO benefits all supply chain participants under integrated channels and performs best when combined with return policies (see also Cachon, 2003).

In related literature, Al-Adwan and Yaseen (2023) studied social commerce and found that lenient return policies significantly reduced seller uncertainty, based on empirical evidence from 471 customers. Such a study demonstrates that return policies play a crucial role in shaping new commerce methods in the omnichannel era.

Regarding the research on customer behavior in omnichannel retailing, Zhang et al. (2020) examined how online-first retailers increasingly adopt physical showrooms in their omnichannel strategy, allowing customers to examine products offline before making online purchases. This creates two types of showrooming behavior: intra- and inter-product showrooming, in which customers inspect one product offline but purchase a different product online. Their research revealed that inter-product showrooming benefits manufacturers with online-exclusive products, while harming those with dual-channel displays. This practice only benefits retailers when showrooming intensity is moderate.

Finally, this study owes Gu and Tayi (2017) for their concept of pseudo-showrooming behavior. Their research assumes that customers inspect products at physical stores before purchasing related but different items online. This customer behavior is influenced by two key factors: fit probability and tolerance levels that customers perceive. These key factors provide valuable insights into customer behavior dynamics, particularly regarding tolerance levels in decision-making.

## 3. The Model

### 3.1 Customer behavior

Consider a retailer selling two substitute products: Products 1 and 2. Product 1 is unfamiliar to customers (a new product), whereas Product 2 is a familiar generic good. Product 1 generates an innate product value of $v_1$, and Product 2 generates $v_2$ ($v_1 > v_2$). Product 2 is sold exclusively online at competitive price $\bar{p}_2$. The retailer sells Product 1 only in a physical store and customers cannot purchase it elsewhere. The customer population is normalized to one.

Product 1 has two types of attributes: those that can be evaluated immediately upon physical inspection, and those that become apparent only through extended use at home. For example, customers can assess color and appearance when examining the product in a physical store, whereas hidden features and long-term appeal become clear only after one or two weeks of daily use. Since Product 1 is available exclusively in physical stores, customers must visit to evaluate the first type of attribute and determine their initial fit. To assess the second type of attribute, customers must use the product for a certain period. These secondary attributes can only be evaluated through a trial option that determines a customer's tolerance level even when the product does not fit the first type of attribute.

When customers visit a physical store to examine Product 1, they determine whether it fits their preferences for the first type of attribute. The probability of fit is $\alpha$ ($0 < \alpha < 1$), which is identical for all customers and common knowledge. If Product 1 fits, the customer purchases it and obtains utility $v_1 - p_1$ (ignoring transportation costs). If Product 1 does not fit, the retailer may persuade the customer to purchase Product 2 online, yielding utility $v_2 - \bar{p}_2$ (ignoring delay costs). In this case, the retailer may offer a trial option for Product 1 (e.g., one or two weeks). Each customer has a tolerance parameter $\beta$, which determines the utility $\beta v_1 - p_1$ during the trial period. This personal parameter $\beta$ is uniformly distributed on $[0, 1)$ and becomes known only during the trial period. We assume that $\alpha$ and $\beta$ have no correlation. If the customer discovers a low $\beta$, they may return Product 1 for a refund but must pay a return cost $rv_1$ ($0 < r < 1$). After returning and buying Product 2, they receive utility $v_2 - \bar{p}_2 - rv_1$.

The customer's decision sequence is: (1) observe $v_1, v_2, p_1, \bar{p}_2$, and $\alpha$, (2) decide whether to visit the store, (3) determine if Product 1 fits, (4) choose whether to use the trial option if it does not fit and the option is offered, and (5) decide whether to return the product after observing $\beta$ during the trial. The retailer can set $p_1$ but not $\bar{p}_2$. This model examines how the retailer can maximize profits by setting the optimal price for Product 1 to induce the desired customer behavior. Assuming that the retailer always offers a trial option to a misfit customer, the game structure is as illustrated in Figure 1. In this game, *Nature* is introduced as a player that determines product fit and $\beta$ for each customer.

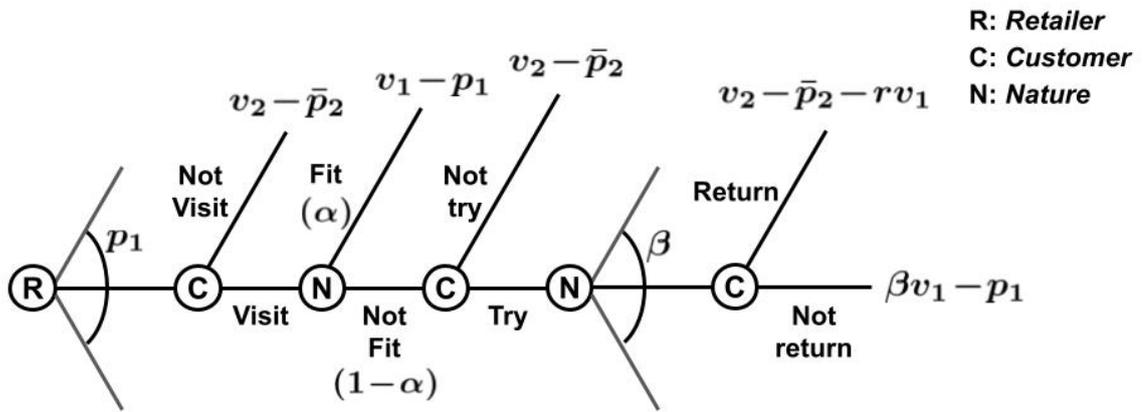

Figure 1. Game Structure.

Before solving the game, let us establish the following assumptions:

*Assumption 1.* All customers are risk neutral.

Under Assumption 1, customers choose the option that maximizes their expected utility before they observe the product fit and their personal parameter $\beta$.

*Assumption 2.* (1) When a customer is indifferent between visiting and not visiting a physical store, they choose to visit; (2) when a customer is indifferent between returning and keeping Product 1, they choose to keep it; and (3) when a customer is indifferent between trying and not trying Product 1, they choose to try it.

Assumption 2 ensures consistent customer behavior in the game solution by specifying that customers behave in the retailer's favor when faced with indifference. The last assumption may involve the retailer's actions: when a customer is undecided about purchasing Product 1 in a physical store, the store clerks must strongly recommend the trial option. Any costs associated with clerks' efforts are ignored.

Table 1. Table of Notation.

| Symbol | Definition |
| --- | --- |
| $v_1$ | Innate product value of Product 1 |
| $v_2$ | Innate product value of Product 2 |
| $p_1$ | Price of Product 1 (endogenous variable) |
| $\bar{p}_2$ | Price of Product 2 (exogenous variable) |
| $\alpha$ | Probability of fit for Product 1 |
| $\beta$ | Tolerance parameter ($0 \leq \beta < 1$) |
| $r$ | Return cost parameter ($0 < r < 1$) |

## 3.2 Solutions of the game

This game can be solved by backward induction. Let us examine two cases: **Case I** where $1/2 \leq r < 1$ and **Case II** where $0 < r < 1/2$.

**Case I ($1/2 \leq r < 1$).**

*Customers' third choice.* First, the customers' decisions to keep or return Product 1 during the trial period are analyzed. If they return it, they pay $rv_1$ and purchase Product 2 from the retailer's online store, yielding utility $v_2 - \bar{p}_2 - rv_1$. If they keep it, they receive utility $\beta v_1 - p_1$ from Product 1. Customers return the product only if $\beta < \bar{\beta}$, where $\bar{\beta} \equiv (p_1 - \bar{p}_2 - rv_1 + v_2)/v_1$. This threshold $\bar{\beta}$ is calculated using the values and prices of both products; a higher $\bar{\beta}$ indicates a greater likelihood of return.

The following lemma describes customer behavior: (Proofs are provided in the Appendix.)

**Lemma 1.** When $p_1 \leq rv_1 - v_2 + \bar{p}_2$, all customers keep the product and receive utility $\beta v_1 - p_1$. When $p_1 \geq (1+r)v_1 - v_2 + \bar{p}_2$, all customers return the product and receive utility $v_2 - \bar{p}_2 - rv_1$. When $rv_1 - v_2 + \bar{p}_2 < p_1 < (1+r)v_1 - v_2 + \bar{p}_2$, customers with $0 \leq \beta < \bar{\beta}$ return the product while others keep it.

According to this lemma, before observing their own $\beta$, customers can expect an expected utility of $(1/2)v_1 - p_1$ in the first case. In the last case, their expected utility is as follows:

$$E_\beta u|_{0<\bar{\beta}<1} \equiv \int_{\bar{\beta}}^1 (\beta v_1 - p_1)\, d\beta + \int_0^{\bar{\beta}} (v_2 - \bar{p}_2 - rv_1)\, d\beta = (1-\bar{\beta})(\tilde{\beta}v_1 - p_1) + \bar{\beta}(v_2 - \bar{p}_2 - rv_1) \quad (1)$$

where $\tilde{\beta} \equiv (1+\bar{\beta})/2$.

*Customers' second choice.* Next, whether customers would try Product 1 with the trial option if it does not fit their preferences is analyzed. The following lemma summarizes the study findings:

**Lemma 2.** Assume $1/2 \leq r < 1$. When $p_1 \leq (1/2)v_1 - v_2 + \bar{p}_2$, customers will try Product 1 when it does not fit their preferences, but will purchase Product 2 without trying Product 1 otherwise.

Based on Lemmas 1 and 2, when Product 1's price is sufficiently low ($p_1 \leq (1/2)v_1 - v_2 + \bar{p}_2$), customers who visit the physical store and find that Product 1 does not fit will try the product and keep it.

*Customers' first choice.* Finally, let us consider whether customers will visit the physical store in the first place, given all observable parameters. Thus, we have the following proposition:

**Proposition 1.** Assume $1/2 \leq r < 1$. When $p_1 \leq (1/2)v_1 - v_2 + \bar{p}_2$, all customers visit the physical store and will try and keep Product 1 even if it does not fit. When $(1/2)v_1 - v_2 + \bar{p}_2 < p_1 \leq v_1 - v_2 + \bar{p}_2$, all customers visit the physical store but will not try Product 1 if it does not fit. When $p_1 > v_1 - v_2 + \bar{p}_2$, no customers visit the physical store; instead, they purchase Product 2.

The results of **Case I**, where the return cost is relatively high ($r \geq 1/2$), are comprehensively presented in Table 2. Table 2 outlines how different price ranges for Product 1 influence customer behavior, ultimate purchase decisions, and the resulting profit outcomes for the retailer. As shown in the detailed breakdown, price thresholds create distinct regions of customer behavior, leading to varying levels of product adoption and profitability.

The last row of the Table 2 shows the retailer's profits. Assuming zero costs for selling products and operating the store, sales are equivalent to profits in this model. To simplify our model, we also exclude salvage values and disposal costs associated with product returns, abstracting away these real-world complexities that influence retailer decisions. When Product 1's price is sufficiently low, all customers buy Product 1 and the profit equals $p_1 \times 1$. At a high price, the profit is $\bar{p}_2 \times 1$. At intermediate prices, a proportion $\alpha$ of customers buy Product 1 while others buy Product 2, resulting in a profit of $\alpha p_1 + (1-\alpha)\bar{p}_2$. Let $\Pi_1$, $\overline{\Pi}_2$, and $\Pi_3$ represent these three profits, respectively:

$$\Pi_1 \equiv p_1, \qquad \overline{\Pi}_2 \equiv \bar{p}_2, \qquad \Pi_3 \equiv \alpha p_1 + (1-\alpha)\bar{p}_2. \tag{2}$$

Table 2. The Results of **Case I** ($1/2 \leq r < 1$).

|  | $p_1 \leq (1/2)v_1 - v_2 + \bar{p}_2$ | $(1/2)v_1 - v_2 + \bar{p}_2 < p_1 \leq v_1 - v_2 + \bar{p}_2$ | $p_1 > v_1 - v_2 + \bar{p}_2$ |
|---|---|---|---|
| Customer Behavior | Visit physical store. Try and keep Product 1 if it does not fit. | Visit physical store. Do not try Product 1 if it does not fit. | Do not visit physical store. |
| Purchase Decision | All customers buy Product 1. | A proportion $\alpha$ buys Product 1; others buy Product 2. | All customers buy Product 2. |
| Retailer's Profit | $p_1 \ (\equiv \Pi_1)$ | $\alpha p_1 + (1-\alpha)\bar{p}_2 \ (\equiv \Pi_3)$ | $\bar{p}_2 \ (\equiv \overline{\Pi}_2)$ |

Figure 2 shows these three profits as functions of the price of Product 1. As illustrated, $\Pi_1$ and $\Pi_3$ take the form of increasing linear curves, while $\overline{\Pi}_2$ appears as a horizontal line. Since $\Pi_3 > \overline{\Pi}_2$ at $p_1 =$ $v_1 - v_2 + \bar{p}_2$, the retailer's maximum profit occurs either at $\Pi_1$ when $p_1 = (1/2)v_1 - v_2 + \bar{p}_2$ (if the price is positive) or at $\Pi_3$ when $p_1 = v_1 - v_2 + \bar{p}_2$.

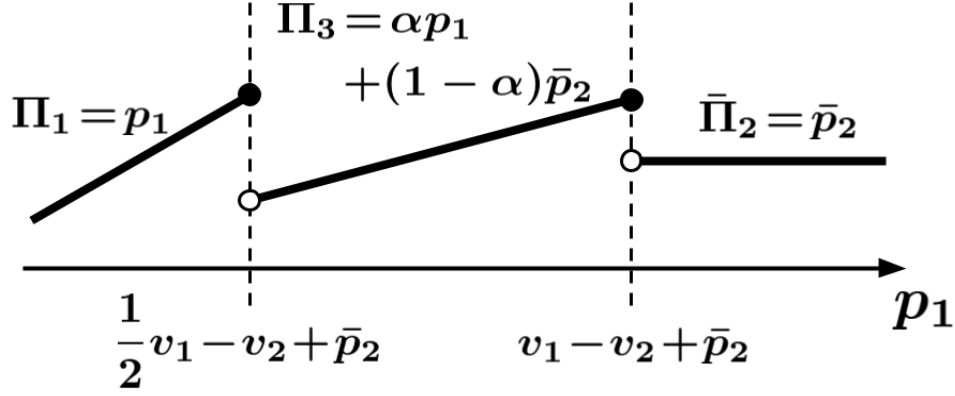

Figure 2. Retailer's Profit (**Case I**).

Determining which profit is larger depends on the parameters $v_1$, $v_2$, and $\alpha$. The following proposition determines the optimal price of Product 1 that maximizes the retailer's profits:

**Proposition 2.** Assume $r \geq 1/2$ and $\alpha \neq 1/2$. When $(v_1/v_2) > (2 - 2\alpha)/(1 - 2\alpha)$, the retailer sets the price of Product 1 to $(1/2)v_1 - v_2 + \bar{p}_2$. At this price, all customers visit the physical store and try and keep Product 1 if it does not fit. When $(v_1/v_2) < (2 - 2\alpha)/(1 - 2\alpha)$, the retailer sets the price of Product 1 to $v_1 - v_2 + \bar{p}_2$. At this price, all customers visit the physical store, but do not try Product 1 if it does not fit.

Figure 3 illustrates the optimal price regions plotted on a plane with axes $\alpha$ and the ratio $v_1/v_2$. The maximum values of $\Pi_1$ and $\Pi_3$ are denoted as $\bar{\Pi}_1$ and $\bar{\Pi}_3$, respectively; the region where $\bar{\Pi}_1 > \bar{\Pi}_3$ appears in the upper-left portion, while the region where $\bar{\Pi}_1 < \bar{\Pi}_3$ appears in the lower-right portion. Based on these results, the retailer is more inclined to offer the trial option when $\alpha$ is smaller and the ratio $v_1/v_2$ is larger.

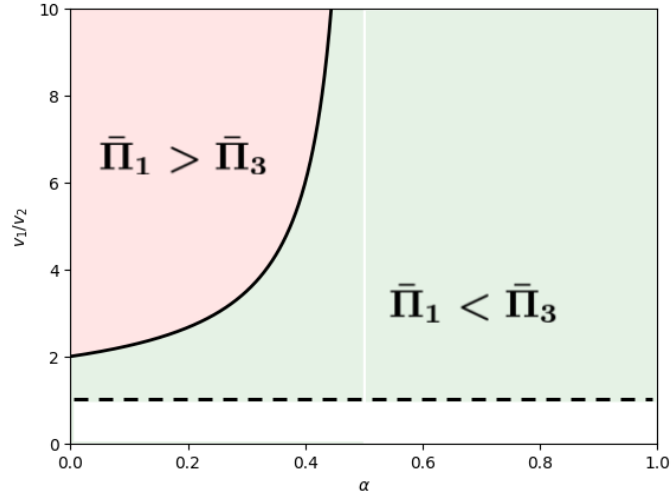

Figure 3. Optimal Price Regions (**Case I**).

Before proceeding with further discussion, let us examine **Case II** where $0 < r < 1/2$.

**Case II ($0 < r < 1/2$).**

*Customers' third choice.* Customers' decisions to keep or return Product 1 during the trial period remain identical to **Case I** because whether $r$ is larger or smaller than one-half does not affect this decision. Therefore, Lemma 1 also applies to this case.

*Customers' second choice.* Next, whether customers try Product 1 with the trial option if it does not fit their preferences is analyzed. Unlike **Case I**, in this case, $rv_1 - v_2 + \bar{p}_2 < (1/2)v_1 - v_2 - \bar{p}_2$, which affects customer behavior regarding Product 1's price. This behavior is described by the following lemma:

**Lemma 3.** Assume $0 < r < 1/2$. When $p_1 \leq (1 + r - \sqrt{2r})v_1 - v_2 + \bar{p}_2$, customers will try Product 1 when it does not fit their preferences, but will purchase Product 2 without trying Product 1 otherwise.

Note that in **Case II** ($0 < r < 1/2$), we have $rv_1 - v_2 + \bar{p}_2 < (1 + r - \sqrt{2r})v_1 - v_2 + \bar{p}_2 < (1 + r)v_1 - v_2 + \bar{p}_2$. Based on Lemmas 1 and 3, all misfit customers will try Product 1 when $p_1 \leq (1 + r - \sqrt{2r})v_1 - v_2 + \bar{p}_2$, and all of them will keep it only if $p_1 \leq rv_1 - v_2 + \bar{p}_2$. If $rv_1 - v_2 + \bar{p}_2 < p_1 \leq (1 + r - \sqrt{2r})v_1 - v_2 + \bar{p}_2$, while some trial customers will return it.

*Customers' first choice.* Finally, let us consider whether customers will visit the physical store in the first place, given all observable parameters. Thus, we have the following proposition for **Case II**:

**Proposition 3.** Assume $0 < r < 1/2$. When $p_1 \leq rv_1 - v_2 + \bar{p}_2$, all customers visit the physical store and will try and keep Product 1 even if it does not fit. When $rv_1 - v_2 + \bar{p}_2 < p_1 \leq (1 + r - \sqrt{2r})v_1 - v_2 + \bar{p}_2$, all customers visit the physical store and try Product 1 if it does not fit; those with $0 < \beta < \bar{\beta}$ will return it, while others will keep it. When $(1 + r - \sqrt{2r})v_1 - v_2 + \bar{p}_2 < p_1 \leq v_1 - v_2 + \bar{p}_2$, all customers visit the physical store but will not try Product 1 if it does not fit. When $p_1 > v_1 - v_2 + \bar{p}_2$, no customers will visit the physical store; instead, they will purchase Product 2.

Table 3 shows the retailer's profit in **Case II**. Unlike **Case I**, we have four different regions for different customer behaviors and purchase decisions. In this case, some customers will try Product 1 if it does not fit but might return it if their own tolerance $\beta$ is sufficiently low.

Table 3. The Results of **Case II** ($0 < r < 1/2$).

|  | $p_1 \leq rv_1 - v_2 + \bar{p}_2$ | $rv_1 - v_2 + \bar{p}_2 < p_1 \leq (1 + r - \sqrt{2r})v_1 - v_2 + \bar{p}_2$ | $(1 + r - \sqrt{2r})v_1 - v_2 + \bar{p}_2 < p_1 \leq v_1 - v_2 + \bar{p}_2$ | $p_1 > v_1 - v_2 + \bar{p}_2$ |
|---|---|---|---|---|
| Customer Behavior | Visit physical store. Try and keep Product 1 if it does not fit. | Visit physical store. Try Product 1 if it does not fit. A proportion $\bar{\beta}$ return it; others keep it. | Visit physical store. Do not try Product 1 if it does not fit. | Do not visit physical store. |
| Purchase Decision | All customers buy Product 1. | A proportion $[\alpha + (1 - \alpha)(1 - \bar{\beta})]$ buy Product 1; others buy Product 2. | A proportion $\alpha$ buy Product 1; others buy Product 2. | All customers buy Product 2. |
| Retailer's Profit | $p_1 \; (\equiv \Pi_1)$ | $[\alpha + (1 - \alpha)(1 - \bar{\beta})]p_1 + (1 - \alpha)(1 - \bar{\beta})\bar{p}_2 \; (\equiv \Pi_4)$ | $\alpha p_1 + (1 - \alpha)\bar{p}_2 \; (\equiv \Pi_3)$ | $\bar{p}_2 \; (\equiv \bar{\Pi}_2)$ |

We denote the retailer's profit in the second region by $\Pi_4$:

$$\Pi_4 \equiv [\alpha + (1 - \alpha)(1 - \bar{\beta})]p_1 + (1 - \alpha)\bar{\beta}\bar{p}_2 = \gamma p_1 + (1 - \gamma)\bar{p}_2. \quad (3)$$

where $\gamma \equiv \alpha + (1 - \alpha)(1 - \bar{\beta})$. This is a quadratic function of $p_1$ which is concave and has a maximum point. The following lemma specifies whether this maximum point falls inside or outside the region, and determines whether the retailer has an interior or corner solution to maximize profit:

**Lemma 4.** Let $p_1^* \equiv \text{argmax}_{p_1} \Pi_4$. (1) $p_1^*$ is larger than $rv_1 - v_2 + \bar{p}_2$; (2) $p_1^*$ is larger than $(1 + r - \sqrt{2r})v_1 - v_2 + \bar{p}_2$ when $\alpha$ and $r$ are sufficiently large and smaller otherwise.

This implies that within this price range for Product 1, the retailer must set the price to either $p_1^*$ or $(1 + r - \sqrt{2r})v_1 - v_2 + \bar{p}_2$ when solving the profit maximization problem.

Figures 4a and 4b illustrate the four profit functions with respect to the price of Product 1. In **Case**

II, $\Pi_4$ appears between the regions of $\Pi_1$ and $\Pi_3$, alongside $\Pi_1$, $\bar{\Pi}_2$, and $\Pi_3$. Since $\Pi_3 > \bar{\Pi}_2$ at $p_1 = v_1 - v_2 + \bar{p}_2$, the retailer's optimal solution must be one of three prices: $p_1 = rv_1 - v_2 + \bar{p}_2$, $p_1 = (1 + r - \sqrt{2r})v_1 - v_2 + \bar{p}_2$, or $p_1 = v_1 - v_2 + \bar{p}_2$. Figure 4a shows the case where $\Pi_4$ is always increasing in the second region, while Figure 4b shows the case with a maximum point within the region.

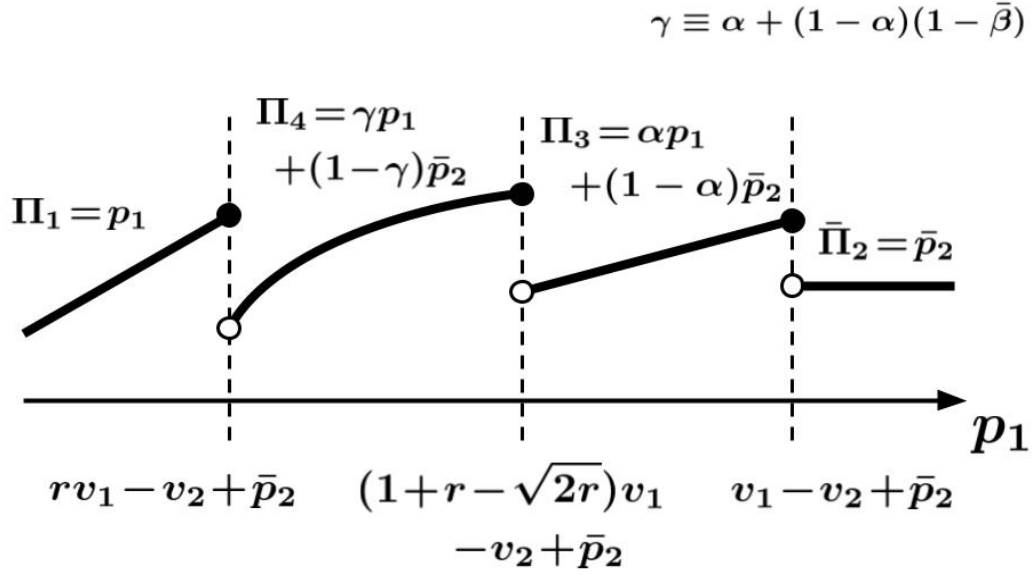

Figure 4a. Retailer's Profit with Corner $\Pi_4$ (**Case II**).

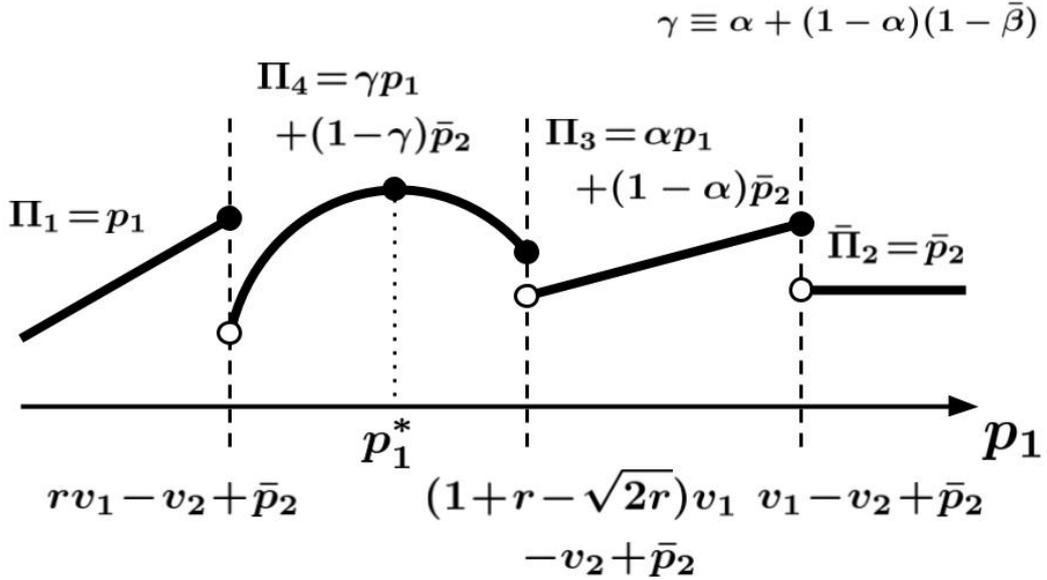

Figure 4b. Retailer's Profit with Interior $\Pi_4$ (**Case II**).

Note that $\Pi_1$ and $\Pi_3$ are increasing in their respective regions, while $\Pi_4$ may have either a corner or interior solution. The following notation is used to represent the maximum values of these profit functions:

$$\begin{aligned}
\overline{\Pi}_1 &\equiv p_1, \quad \text{where } p_1 = rv_1 - v_2 + \bar{p}_2, \\
\overline{\Pi}_3 &\equiv \alpha p_1 + (1-\alpha)\bar{p}_2, \quad \text{where } p_1 = v_1 - v_2 + \bar{p}_2, \\
\overline{\Pi}_4^C &\equiv \gamma p_1 + (1-\gamma)\bar{p}_2, \quad \text{where } p_1 = \left(1 + r - \sqrt{2r}\right)v_1 - v_2 + \bar{p}_2, \\
\overline{\Pi}_4^I &\equiv \gamma p_1 + (1-\gamma)\bar{p}_2, \quad \text{where } p_1 = p_1^*.
\end{aligned} \quad (4)$$

The following proposition summarizes the findings for **Case II**: (The proof is omitted.)

**Proposition 4.** Assume $0 < r < 1/2$. The retailer's optimal pricing strategy is as follows:

(1) If $p_1^* > \left(1 + r - \sqrt{2r}\right)v_1 - v_2 + \bar{p}_2$;

When $\overline{\Pi}_1 > \overline{\Pi}_3$ and $\overline{\Pi}_1 > \overline{\Pi}_4^C$, set Product 1's price to $rv_1 - v_2 + \bar{p}_2$. This results in all customers visiting the physical store, trying Product 1, and keeping it, even if it does not fit. When $\overline{\Pi}_3 > \overline{\Pi}_1$ and $\overline{\Pi}_3 > \overline{\Pi}_4^C$, set Product 1's price to $v_1 - v_2 + \bar{p}_2$. This leads to all customers visiting the physical store but not trying Product 1 if it does not fit. When $\overline{\Pi}_4^C > \overline{\Pi}_1$ and $\overline{\Pi}_4^C > \overline{\Pi}_3$, set Product 1's price to $\left(1 + r - \sqrt{2r}\right)v_1 - v_2 + \bar{p}_2$. This results in all customers visiting the physical store and trying Product 1 if it does not fit; customers with $0 < \beta < \bar{\beta}$ will return it, while others will keep it.

(2) If $p_1^* < \left(1 + r - \sqrt{2r}\right)v_1 - v_2 + \bar{p}_2$;

When $\overline{\Pi}_1 > \overline{\Pi}_3$ and $\overline{\Pi}_1 > \overline{\Pi}_4^I$, set Product 1's price to $rv_1 - v_2 + \bar{p}_2$. This results in all customers visiting the physical store, trying Product 1, and keeping it, even if it does not fit. When $\overline{\Pi}_3 > \overline{\Pi}_1$ and $\overline{\Pi}_3 > \overline{\Pi}_4^I$, set Product 1's price to $v_1 - v_2 + \bar{p}_2$. This leads to all customers visiting the physical store but not trying Product 1 if it does not fit. When $\overline{\Pi}_4^I > \overline{\Pi}_1$ and $\overline{\Pi}_4^I > \overline{\Pi}_3$, set Product 1's price to $p_1^*$. This results in all customers visiting the physical store and trying Product 1 if it does not fit; customers with $0 < \beta < \bar{\beta}$ will return it, while others will keep it.

## 4. Discussion

### 4.1 Implication of the results

Let us focus on the implications of the game analysis results. The threshold $\bar{\beta}$ plays a crucial role; customers will return Product 1 if they observe a low tolerance parameter $\beta$ such that $0 < \beta < \bar{\beta}$. Since $\partial \bar{\beta}/\partial p_1 = 1/v_1 > 0$, the retailer can induce customers to keep Product 1 by sacrificing increased sales. However, if the innate value of Product 1, $v_1$, is sufficiently low, the foregone sales from inducing keeping will become costly. In this case, the retailer must abandon their keeping strategy and accept that customers will not try Product 1 if they dislike it.

Figure 5 illustrates the regions for the retailer's optimal strategy on the planes of $\alpha$ and $r$. Figure 5a shows the case where $v_1 = 2, v_2 = 1, \bar{p}_2 = 0$. In **Case I** ($1/2 \leq r < 1$), $\bar{\Pi}_3$ is the maximum profit for any $\alpha$ and $r$. In **Case II** ($0 < r < 1/2$), $\bar{\Pi}_3$ is the maximum profit in Region 3, while $\bar{\Pi}_4^C$ and $\bar{\Pi}_4^I$ are the maximum profits in Regions 4(C) and 4(I), respectively. No region exists where $\bar{\Pi}_1$ yields the maximum profit for any case. Figure 5b illustrates the case where $v_1 = 3, v_2 = 1, \bar{p}_2 = 0$. In **Case I** ($1/2 \leq r < 1$), Region 1 emerges as the area where $\bar{\Pi}_1$ yields the maximum profit. Figure 5c ($v_1 = 10, v_2 = 1, \bar{p}_2 = 0$) illustrates the case in which the value of Product 1 is substantially higher than that of Product 2. Under these conditions, Region 3 contracts, whereas Regions 4(I), 4(C), and 1 expand.

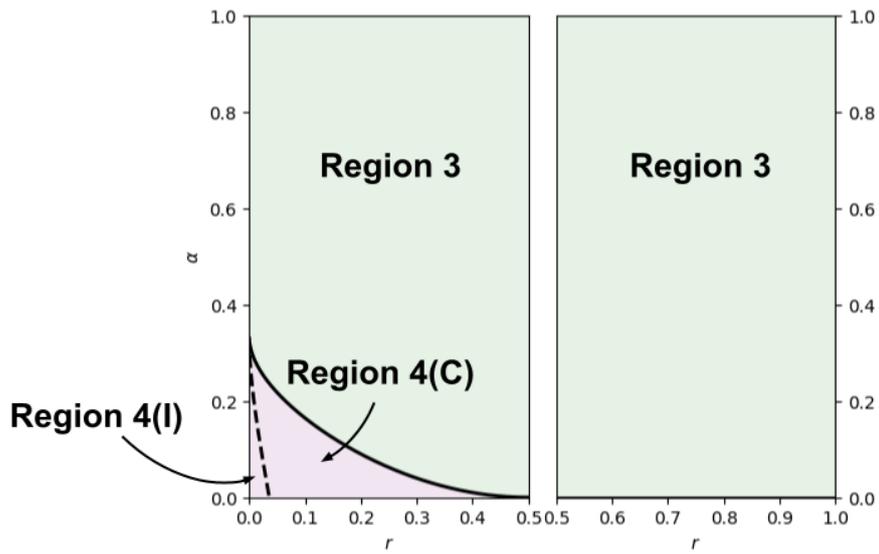

Figure 5a. Regions of Maximum Profit ($v_1 = 2, v_2 = 1, \bar{p}_2 = 0$).

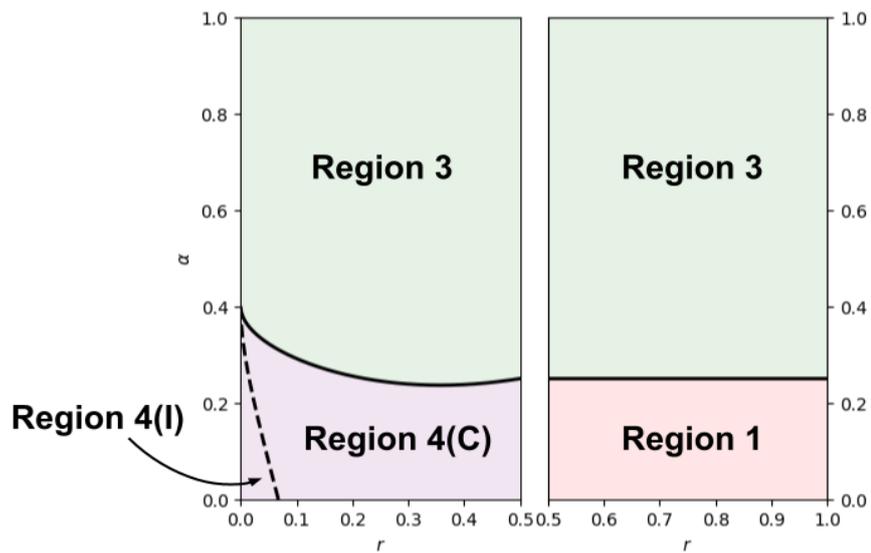

Figure 5b. Regions of Maximum Profit ($v_1 = 3, v_2 = 1, \bar{p}_2 = 0$).

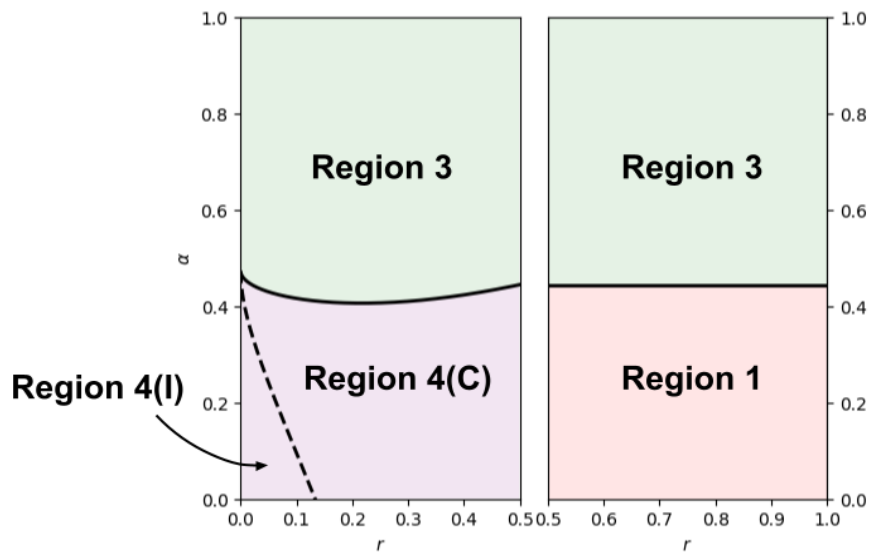

Figure 5c. Regions of Maximum Profit ($v_1 = 10, v_2 = 1, \bar{p}_2 = 0$).

Based on these numerical examples and Propositions 2 and 4, the following conclusion can be drawn:

**Argument 1.** When Product 1's value sufficiently exceeds that of Product 2, misfit consumers are more likely to keep Product 1 after trying it. However, when the probability of fit ($\alpha$) is higher, misfit consumers are less likely to try the product at all.

### 4.2 Advantages of having two channels

Another interesting finding is that a pricing strategy that induces *all* customers to *try and keep* a misfit product might be suboptimal. In **Case II** ($0 < r < 1/2$), Region 1 ($p_1 < rv_1 - v_2 + \bar{p}_2$) cannot be considered the optimal price region for any $\alpha$ and $r$. This region may not even exist when $rv_1 - v_2 + \bar{p}_2 < 0$. When the return cost is low, the retailer may encourage all customers to try the product, but allow some returns. In this scenario, the retailer must rely on the online channel's sales of Product 2 as a profit source when using the product to attract customers who find Product 1 unsuitable. This necessitates the maintenance of two distinct channels: a physical store for product trials, and an online store for alternative options when products do not fit. The model demonstrates how retailers can coordinate the two channels through strategic product pricing in omnichannel systems. Thus, the following assertion can be made:

**Argument 2.** In an omnichannel system, both physical and online channels are essential when it is not optimal to have a pricing strategy in which *all* customers *try and keep* products regardless of fit. Instead, retailers should maintain two complementary channels: physical stores for product trials and online stores to provide alternatives when products do not fit.

### 4.3 The role of store clerks in the omnichannel system

Based on Assumption 2(3), when customers are indifferent between trying and not trying Product 1, they choose to try it. This allows the retailer to set the price to $(1/2)v_1 - v_2 + \bar{p}_2$ in **Case I** and $(1 + r - \sqrt{2r})v_1 - v_2 + \bar{p}_2$ in **Case II** if these are optimal solutions. At these prices, customers will choose to try Product 1. However, without this assumption, customers may opt not to try the product because the expected utility is equal. This is where store clerks play a crucial role: they can persuade customers to try Product 1 in person, thereby securing $\bar{\Pi}_1$ and $\bar{\Pi}_4^C$ for **Cases I** and **II**, respectively. Since this kind of persuasion is not possible online, the following assertion can be made:

**Argument 3.** In an omnichannel system, store clerks provide a unique advantage by encouraging customers to try products that may not initially match their preferences. This represents a key benefit of establishing an omnichannel strategy.

It is worth reconsidering the omnichannel strategy by emphasizing the role of customer persuasion. Customer persuasion represents a unique opportunity for physical stores to leverage their human touchpoints in ways online channels cannot replicate. The strategic advantage of customer interaction and persuasion capabilities should be carefully evaluated and integrated into a broader omnichannel framework. The model identifies situations in which customers are ambivalent about purchasing a product while at the store. Without persuasion, customers may forget about the product and choose to find alternatives online. This decision is crucial for the retailer, because potential profits may be lost. This implies that retailers must establish physical stores with expert clerks to encourage customers to try the products and reassess their value.

## 4.4 The possibility of retailer-paid return fees

Finally, the question of whether retailers should cover return costs when customers try and return a misfit product after the trial period is examined. If a retailer announces full coverage of return costs, customers know that they will not pay for returns during the trial period, which encourages them to try the product. In the model, this coverage policy does not affect customer behavior or retailer profits in **Case I**. In **Case II**, Regions 1 and 3 disappear when $rv_1 - v_2 + \bar{p}_2 < 0$ and $(1 + r - \sqrt{2r})v_1 - v_2 + \bar{p}_2 = v_1 - v_2 + \bar{p}_2$ when $r = 0$. If $\Pi_4$ lacks an interior solution, the retailer must price Product 1 at $v_1 - v_2 + \bar{p}_2$, leading all misfit customers to return it and buy Product 2, resulting in a profit of $\bar{\Pi}_3 - rv_1$. In this case, there is no incentive to offer returns coverage. However, if $\Pi_4$ has an interior solution, the retailer can set $p_1 < v_1 - v_2 + \bar{p}_2$, causing some customers to keep Product 1 despite the misfit. When $\bar{\Pi}_4^I - rv_1$ is sufficiently large, a return coverage policy may be profitable for the retailer. Thus, the following assertion can be made:

**Argument 4.** When return costs are sufficiently high, retailers have no incentive to implement a coverage policy because it affects neither customer behavior nor profits. However, when return costs are low, offering return coverage may prove profitable for retailers.

## 5. Conclusion

In this study, research questions about trial and return options in omnichannel systems were proposed, a model was formulated using an extensive game structure, and the arguments addressing these research questions were discussed. By examining optimal customer behavior, the rationale for establishing an omnichannel system that integrates physical and online stores in terms of trial and return options was confirmed. This section addresses some of the limitations of the model with regard to the results and insights.

First, some realistic complex settings were abstracted to focus solely on strategic customer behavior, independent of the model parameters. Travel and hassle costs, which are the expenses incurred by customers when visiting physical stores, were ignored. Similarly, salvage discounts, which represent losses from product returns, were also ignored.

Second, the various channels and mechanisms through which customers could return products that do not meet their needs were not explored. This includes in-person returns to physical retail locations and remote returns via postal or courier services. The choice of return method could potentially impact customer behavior and retailer costs in ways that warrant further investigation in future research.

Third, the model relies on a key simplifying assumption regarding market dynamics: Product 2 is positioned in a perfectly competitive market environment, which consequently removes the retailer's ability to exercise pricing control over this product. This deliberate constraint in the model structure allowed for a specific analytical focus on the pricing strategy development for Product 1, a product whose characteristics and value proposition remain initially uncertain to potential customers. This assumption, while helpful for model tractability, may not fully capture the complexity of real-world market conditions, where retailers often have some degree of pricing power across their entire product portfolio.

Due to this study's primary focus on consumer behavior and decision-making patterns, we employed a simplified theoretical model that emphasizes key behavioral aspects while abstracting from certain real-world complexities. While this approach allowed us to derive clear analytical insights about customer choices and retailer strategies, future research can address these limitations by incorporating additional market factors, expanding the model's scope, and considering more nuanced aspects of omnichannel retail operations.

**Appendix**

*Proof of Lemma 1.* When $p_1 \leq rv_1 - v_2 + \bar{p}_2$ (equivalently, when $\bar{\beta} \leq 0$), for any customer, $\bar{\beta} \leq \beta < 1$, and from Assumption 2 (1), all customers keep Product 1 because $\beta v_1 - p_1 \geq v_2 - \bar{p}_2 - rv_1$. When $p_1 \geq (1+r)v_1 - v_2 + \bar{p}_2$ (equivalently, $\bar{\beta} \geq 1$), for any customer, $0 < \beta < \bar{\beta}$, and all customers return Product 1 because $v_2 - \bar{p}_2 - rv_1 > \beta v_1 - p_1$. When $rv_1 - v_2 + \bar{p}_2 < p_1 < (1+r)v_1 - v_2 + \bar{p}_2$ (equivalently, when $0 < \bar{\beta} < 1$), customers return Product 1 if $0 < \beta < \bar{\beta}$ and keep it otherwise.

*Proof of Lemma 2.* Three cases were considered based on the threshold value $\bar{\beta}$: when $\bar{\beta} \leq 0$, when $\bar{\beta} \geq 1$, and when $0 < \bar{\beta} < 1$. These cases are equivalent to $p_1 \leq rv_1 - v_2 + \bar{p}_2$, $p_1 \geq (1+r)v_1 - v_2 + \bar{p}_2$, and $rv_1 - v_2 + \bar{p}_2 < p_1 < (1+r)v_1 - v_2 + \bar{p}_2$, respectively. In the first case, customers receive the expected utility $(1/2)v_1 - p_1$ if trying and $v_2 - \bar{p}_2$ otherwise. Therefore, if $p_1 \leq (1/2)v_1 - v_2 + \bar{p}_2$, they will try. In the second case, all customers will return and pay $rv_1$, which means that they will not try and will purchase Product 2 instead. In the last case, they will have expected utility $E_\beta u|_{0<\bar{\beta}<1}$ with trial but $v_2 - \bar{p}_2$ without trial. Considering $E_\beta u|_{0<\bar{\beta}<1}$ as a function of $p_1$, it is a convex function with the minimum point at $(1+r)v_1 - v_2 + \bar{p}_2$. Denoting $\Delta(p_1) \equiv E_\beta u|_{0<\bar{\beta}<1} - (v_2 - \bar{p}_2)$, $\Delta(\cdot)$ is decreasing when $p_1 < (1+r)v_1 - v_2 + \bar{p}_2$. Let $p_1^A \equiv rv_1 - v_2 + \bar{p}_2$ and $p_1^B \equiv (1+r-\sqrt{2r})v_1 - v_2 + \bar{p}_2$. Given that $E_\beta u|_{0<\bar{\beta}<1} = (v_2 - \bar{p}_2)$ at $p_1^B$ and $p_1^B \leq p_1^A$ under the case of $r \geq 1/2$, for any $p_1 \in (rv_1 - v_2 + \bar{p}_2, (1+r)v_1 - v_2 + \bar{p}_2)$,

$$0 = \Delta(p_1^B) \geq \Delta(p_1^A) > \Delta(p_1). \tag{A1}$$

Therefore, $E_\beta u|_{0<\bar{\beta}<1} < (v_2 - \bar{p}_2)$ at any value of $p_1$ in the last case, indicating that customers will not attempt to try Product 1.

*Proof of Proposition 1.* When $p_1 \leq (1/2)v_1 - v_2 + \bar{p}_2$, customers who visit the physical store receive utility $v_1 - p_1$ with probability $\alpha$ and expected utility $(1/2)v_1 - p_1$ with probability $(1 - \alpha)$, based on Lemma 2. If they do not visit, they receive utility $v_2 - \bar{p}_2$ by purchasing Product 2. Since $\alpha(v_1 - p_1) + (1 - \alpha)((1/2)v_1 - p_1) > v_2 - p_2$, all customers visit the physical store. When $p_1 > (1/2)v_1 - v_2 + \bar{p}_2$, customers who visit the physical store receive utility $v_1 - p_1$ with probability $\alpha$ and utility $v_2 - \bar{p}_2$ with probability $(1 - \alpha)$, based on Lemma 2. If $p_1 < v_1 - v_2 + \bar{p}_2$, then $\alpha(v_1 - p_1) + (1 - \alpha)(v_2 - \bar{p}_2) > v_2 - p_2$, meaning all customers visit; otherwise, no customers visit. When $1/2 \leq r < 1$, we have $(1/2)v_1 - v_2 + \bar{p}_2 \leq rv_1 - v_2 + \bar{p}_2 < v_1 - v_2 + \bar{p}_2$, which completes the proof.

*Proof of Proposition 2.* Assume $r > 1/2$ and $\alpha \neq 1/2$. If $(v_1/v_2) > (2 - 2\alpha)/(1 - 2\alpha)$, then $(1/2)v_1 - v_2 + \bar{p}_2 > \alpha(v_1 - v_2 + \bar{p}_2) + (1 - \alpha)\bar{p}_2$. The retailer should set the price of Product 1 to $(1/2)v_1 - v_2 + \bar{p}_2$ as long as it is positive, and all customers should visit the physical store and try to keep Product 1 if it does not fit. For any $\alpha \in (0,1)$, $(1/2)v_1 - v_2 + \bar{p}_2 > 0$ when $(v_1/v_2) > (2 - 2\alpha)/(1 - 2\alpha)$, where the price is feasible. If $(v_1/v_2) < (2 - 2\alpha)/(1 - 2\alpha)$, then $(1/2)v_1 - v_2 + \bar{p}_2 < \alpha(v_1 - v_2 + \bar{p}_2) + (1 - \alpha)\bar{p}_2$. The retailer should set the price of Product 1 to $v_1 - v_2 + \bar{p}_2$ and all customers should visit the physical store but not try Product 1 if it does not fit. The price $v_1 - v_2 + \bar{p}_2$ is always positive, based on the assumption that $v_1 > v_2$.

*Proof of Lemma 3.* Three cases are considered: $p_1 \leq rv_1 - v_2 + \bar{p}_2$, $p_1 \geq (1 + r)v_1 - v_2 + \bar{p}_2$, and $rv_1 - v_2 + \bar{p}_2 < p_1 < (1 + r)v_1 - v_2 + \bar{p}_2$. In the first case, customers receive the expected utility $(1/2)v_1 - p_1$ if trying and $v_2 - \bar{p}_2$ otherwise. As $p_1 < rv_1 - v_2 + \bar{p}_2 < (1/2)v_1 - v_2 + \bar{p}_2$, all customers will try in this case. In the second case, all customers will return and pay $rv_1$ for returns, which means that they will not try and will purchase Product 2 instead. In the final case, they will have expected utility $E_\beta u|_{0<\bar{\beta}<1}$ with trial but $v_2 - \bar{p}_2$ without trial. Considering $E_\beta u|_{0<\bar{\beta}<1}$ as a function of $p_1$, it is a convex function with the minimum point at $(1 + r)v_1 - v_2 + \bar{p}_2$. Denoting $\Delta(p_1) \equiv E_\beta u|_{0<\bar{\beta}<1} - (v_2 - \bar{p}_2)$, $\Delta(\cdot)$ is decreasing when $p_1 < (1 + r)v_1 - v_2 + \bar{p}_2$. Let $p_1^A \equiv rv_1 - v_2 +$

$\bar{p}_2$ and $p_1^B \equiv (1 + r - \sqrt{2r})v_1 - v_2 + \bar{p}_2$. Given that $E_\beta u|_{0<\bar{\beta}<1} = (v_2 - \bar{p}_2)$ at $p_1^B$ and $p_1^B > p_1^A$ under the case of $r < 1/2$,

$$\Delta(p_1^A) > \Delta(p_1^B) = 0. \tag{A2}$$

This means that $p_1^B \in (rv_1 - v_2 + \bar{p}_2, (1 + r)v_1 - v_2 + \bar{p}_2)$ such that $\Delta(p_1^B) = 0$ because $p_1^B < (1 + r)v_1 - v_2 + \bar{p}_2$. Therefore, if $p_1 < (1 + r - \sqrt{2r})v_1 - v_2 + \bar{p}_2$, they will make an attempt.

*Proof of Proposition 3.* When $p_1 \leq rv_1 - v_2 + \bar{p}_2$, customers who visit the physical store receive utility $v_1 - p_1$ with probability $\alpha$ and expected utility $(1/2)v_1 - p_1$ with probability $(1 - \alpha)$, based on Lemma 3. If they do not visit, they receive utility $v_2 - \bar{p}_2$ by purchasing Product 2. Since $\alpha(v_1 - p_1) + (1 - \alpha)((1/2)v_1 - p_1) \geq v_2 - \bar{p}_2$, all customers visit the physical store. When $rv_1 - v_2 + \bar{p}_2 < p_1 < (1 + r - \sqrt{2r})v_1 - v_2 + \bar{p}_2$, customers who visit the physical store receive utility $v_1 - p_1$ with probability $\alpha$ and utility $E_\beta u|_{0<\bar{\beta}<1}$ with probability $(1 - \alpha)$, based on Lemma 3. If $p_1 \leq (1 + r - \sqrt{2r})v_1 - v_2 + \bar{p}_2$, then $\alpha E_\beta u|_{0<\bar{\beta}<1} + (1 - \alpha)(v_2 - \bar{p}_2) \geq v_2 - \bar{p}_2$, meaning all customers visit. When $p_1 > (1 + r - \sqrt{2r})v_1 - v_2 + \bar{p}_2$, customers who visit the physical store receive utility $v_1 - p_1$ with probability $\alpha$ and utility $v_2 - \bar{p}_2$ with probability $(1 - \alpha)$, based on Lemma 3. If $p_1 \leq v_1 - v_2 + \bar{p}_2$, then $\alpha(v_1 - p_1) + (1 - \alpha)(v_2 - \bar{p}_2) \geq v_2 - \bar{p}_2$, meaning all customers visit; otherwise, no customers visit. When $0 < r < 1/2$, we have $rv_1 - v_2 + \bar{p}_2 < (1 + r - \sqrt{2r})v_1 - v_2 + \bar{p}_2 < v_1 - v_2 + \bar{p}_2$, which completes the proof.

*Proof of Lemma 4.* The unique maximal point of $\Pi_4$ is as follows:

$$p_1^* = \frac{1}{2(1 - \alpha)}\{\alpha v_1 + (1 - \alpha)[(1 + r)v_1 - v_2 + 2\bar{p}_2]\}. \tag{A3}$$

(1) This can be proven from the following:

$$p_1^* - (rv_1 - v_2 + \bar{p}_2) = \frac{1}{2(1-\alpha)}\{\alpha v_1 + [(1-r)v_1 + v_2]\} > 0. \tag{A4}$$

(2) We have the following relationship:

$$p_1^* - \left((1+r-\sqrt{2r})v_1 - v_2 + \bar{p}_2\right) = \frac{1}{2(1-\alpha)}\{[(2\sqrt{2r}-r-1)v_1 + v_2] + \alpha[(2\sqrt{2r}-r-2)v_1 + v_2]\}. \tag{A5}$$

Let us denote this by $\Delta_4(\alpha, r)$ as a function of $\alpha$ and $r$. We can find that $\Delta_4(0,0) < 0$ and $\Delta_4(1, 1/2) > 0$, and that $\partial\Delta_4/\partial\alpha > 0$ and $\partial\Delta_4/\partial r > 0$, which proves the second part of the lemma.